\documentclass{elsart}

\usepackage{amsmath}
\usepackage{amssymb}

\begin{document}

\begin{frontmatter}

\title{On the measurements regarding random observables}

\author{S. Dumitru}
\address{Department of Physics, Transilvania University, B-dul Eroilor
  29, 
R-2200, Brasov, Romania}
\ead{s.dumitru@unitbv.ro}

\author{A. Boer}
\address{Department of Physics, Transilvania University, B-dul Eroilor
  29, 
R-2200, Brasov, Romania}
\ead{boera@unitbv.ro}

\begin{abstract}
Both classical and respectively quantum observables can be modeled as somewhat similar examples of random variables. In such a model the associated measurements preserve the values spectrum of an observable but change the corresponding probabilistic weights (probability density or respectively the wave function). Such a model ensures theoretical estimations for predicted errors specific to the mean values as well as to the fluctuations of both types of observables. The model stands out of ourdays prevalent opinion that measurement theories must be depicted in entirely different manners in the classical respectively quantum cases.
\end{abstract}

\end{frontmatter}

\section{Introduction}
\label{intro}
According to the modern physics  terminology the term \textit{observable} imply the following two features: (i) it is  a quantity which characterize quantitatively an intrinsic property of a physical system and (ii) it can be evaluated by extrinsic experimental devices through adequate measurement process. For classical (non-quantum) systems, the two features are treated theoretically as clearly separate subjects (The first feature is investigated in the framework of various chapters of classical physics while the second one is discussed in the so called (measuring) error theory). In the case of   quantum systems the theoretical descriptions of the mentioned features are often regarded as fundamentally inseparable  things (see the references \cite{1,2,3} and quoted bibliographies). So in the ourdays publications prevails the opinion that  measurement theories must be depicted in entirely different manners in the classical respectively quantum cases.

In this paper we try to develop (at least partially) a suggestion which reconsiders and removes  the actually predominant opinions regarding the differences between the approaches of classical and quantum measurements. Our suggestion is builded on the idea that, by means of a few minimal settings, both classical and quantum observables can be described mathematically as random variables. In such a description  a measurement preserves the spectrum of values of an observable  but change the corresponding probabilistic weights (probability density or the wave function in classical respectively quantum cases). Within the announced  suggestion the theoretical description of   measurements offer concomitantly  evaluations for measuring changes of mean (expected) values as well as of fluctuations (deviations from the mean) which characterize the observables.

The same suggestion allows to present few unconventional (and natural) considerations about some questions regarding foundations and interpretation of quantum mechanics.

\section{The case of classical observables}
\label{sec:2}
Mathematically, the observables from classical (non-quantum) physics are variables  of both deterministic and random types. The deterministic observables are encountered in newtonian mechanics, simple-thermodynamics (where the fluctuations are neglected) and non-stochastic-electrodynamics. Such an observable  is characterized by an unique value in a given state of the considered system. The observables of random type are specific in probabilistic-thermodynamics (known as phenomenological theory of fluctuations) and in statistical mechanics. For a given system in a well specified state, a random  observable is characterized not by an unique value but by a spectrum (number) of values associated with corresponding probabilities.

We start our announced suggestion by a first setting in which both above mentioned types of classical observables can be described mathematically as being of a single sort that of random variables. The respective things can be presented as follows. Let us consider firstly  a single observable $A$ which by its nature is a true random variable. Such a kind of observable one finds in cases of probabilistic-thermodynamics \cite{4}(e.g. the temperature $T$) respectively of statistical physics \cite{5}(e.g. the modulus $v$ of  a gas molecule velocity). The respective  random variable will be denoted here by the symbol $\mathcal{A}$. As it is known, $\mathcal{A}$ is characterized by a continuous spectrum of values $\Omega _{A} $ (e.g. $\Omega _{A} \Leftrightarrow a\in \left(a_{1}, a_{2} \right)$ if $a$ signifies an individual value of $\mathcal{A}$ ) respectively by the corresponding probability density $w\left(a\right)$. An observable $B$ of deterministic type (such are the ones encountered in newtonian mechanics or in simple-thermodynamics) can be described formally as a random variable 
$\mathcal{B}$ by means of the following reasonable convention. If due of its nature, for a physical system in a given state, $B$ is characterized by a unique value $b_{0}$ then the corresponding random variable $\mathcal{B}$ is endowed with the probability density $w\left(b\right)=\delta \left(b-b_{0} \right)$ (where $\delta \left(\xi \right)$ denote the Dirac's $\delta$ function of argument $\xi$). Associated with the mentioned $w\left(b\right)$ the variable $\mathcal{B}$ has a continuous spectrum of values $\Omega _{B} $ (e.g. $\Omega _{B} \Leftrightarrow b\in \left(b_{1}, b_{2} \right)$ if $b$ denotes an  individual value of $\mathcal{B}$ and $b_{1} \le b_{0} \le b_{2} $). 

A second piece of the announced settings regards the ensemble of observables concerning to the same system in a given state. As it is known, mathematically, the respective observables appear as variables which are not independent among them. That is because due to physical reasons the observables from such an ensemble are incorporated in some relationships of mutual dependence. Such a fact allows to select fom the mentioned ensemble a lot of independent variables while the remaining ones to be regarded as derived (dependent/subsequent) quantities. 

In order to simplify our further discussions we will choose a generic situation. We consider  a system in a given state to be characterized by  a lot of two independent random variables $\mathcal{X}$ and $\mathcal{Y}$ which are endowed with the spectra $\Omega _{X} $ and $\Omega _{Y} $ respectively by the joint distribution of probability $w\left(x,y\right)$. For the same system and state we refer to a set of two derived random quantities $\mathcal{A}$ and $\mathcal{B}$. 
Such a choosing imply the existence of two functional relationships of the form
\begin{equation} \label{eq:1} 
\mathcal{A}=f\left(\mathcal{X},\mathcal{Y}\right)\, ,\quad \mathcal{B}=g\left(\mathcal{X},\mathcal{Y}\right) 
\end{equation} 
These relationships show that, for a given lot $x$ and $y$ of individual values for the independent variables $\mathcal{X}$ and $\mathcal{Y}$, the system is characterized by the set $a=f\left(x,y\right)$ and 
$b=g\left(x,y\right)$ of individual realizations of the dependent observables $\mathcal{A}$ and $\mathcal{B}$. The whole ensembles of values for $a$ and $b$ give the spectra $\Omega_{A}$ and $\Omega_{B}$ of the random observables $\mathcal{A}$ and $\mathcal{B}$.  

In order to compress the discussions we will denote all the observables under attention with the symbols $\mathcal{Z}_{\alpha}$  where 
\begin{equation} \label{eq:2} 
\alpha =\left\{1,2,3,4\right\}\, ,\quad \mathcal{Z}_{\alpha } \in \left\{\mathcal{Z}_{1}, \mathcal{Z}_{2}, \mathcal{Z}_{3}, \mathcal{Z}_{4} \right\}\equiv \left\{\mathcal{A}, \mathcal{B}, \mathcal{X}, \mathcal{Y}\right\} 
\end{equation} 
So we can say  that the main characteristics of the considered system are given by the random observables $\mathcal{Z}_{\alpha}$, by their spectra of values $\Omega_{Z_{\alpha}}$ and by the joint distribution of probability $w\left(x,y\right)$.

By using the above presented settings our suggestion regarding the description of measurements for classical random observables can be modelled as follows. Any measurement, independently of its technical details, can be assumed that it does not change the spectra $\Omega _{Z_{\alpha}}$ of the previously presented random variables $\mathcal{Z}_{\alpha }$. But, if one takes into account the inherent imperfections of the experimental devices, it is credible and rightful the idea that the same measurement must be described as a process which change the probability distribution $w\left(x,y\right)$. Some simplified versions of the respective idea were discussed in few of our previous works 
\cite{6,7,8,9,10}.  

Through the alluded idea and its corresponding developments in details we think that our old and present attempts enlarge and add new elements to some other foregoing approaches 
\cite{11,12,13,14,15,16,17} regarding the theoretical description of measurements.

The above mentioned change of $w\left(x,y\right)$ during a measuring process can schematized as follows. In such a process the \textit{input (in)} information regarding the intrinsic properties of the measured system is converted in \textit{output (out)} information incorporated within the data received on a device recorder. That is why a measurement appears as an \textit{information transmission process}. For the distribution $w\left(x,y\right)$ the respective process evidences two variants denoted here by $w_{in} \left(x,y\right)$ and respectively $w_{out} \left(x,y\right)$. On the one hand the distribution $w_{in} \left(x,y\right)$ describes the intrinsic properties of the measured system. On the other hands the distribution $w_{out} \left(x,y\right)$ incorporates the information about the same system, but obtained on the recorder of measuring device. 

Then, in terms of the above explanations, a measurement regarding the considered system can be modelled theoretically through a transformation of the form
\begin{equation} \label{eq:3} 
w_{in} \left(x,y\right)\to w_{out} \left(x,y\right) 
\end{equation} 
The concrete analytical expression of this transformation requires justifications by taking into account some of the most general characteristics regarding the measuring devices. Among such characteristics of first interest are the following properties (\textbf{\textit{P}}):     
\begin{itemize}
\item[\textbf{\textit{P1:}}] \textit{A good measuring device is stationary in time, i.e. its performances have a sufficiently long standing viability.}
\item[\textbf{\textit{P2:}}] \textit{The same device is forced to  guarantee~a linear superposition of the input signals in giving acceptable output records.}
\end{itemize}  
These properties can be incorporated naturally into the alluded modelling of  a measurement if  the transformation \eqref{eq:3} is written as follows 
\begin{equation} \label{eq:4} 
w_{out} \left(x,y\right)=\int _{\Omega _{X'} }\int _{\Omega _{Y'} } G\left(x,y|x',y'\right)\cdot w_{in} \left(x',y'\right)  \cdot dx'\cdot dy' \end{equation} 
Regarded from the physics perspective the term $G\left(x,y|x',y'\right)$ incorporates the theoretical description of all the characteristics of the measuring device. For an ideal device which ensure 
$w_{out} \left(x,y\right)\equiv w_{in} \left(x,y\right)$ the mentioned term must be of the form
$G\left(x,y|x',y'\right)=\delta \left(x-x'\right)\cdot \delta \left(y-y'\right)$ (with 
$\delta \left(\xi \right)$ denoting the Dirac's function of argument $\xi $).

On the other hand, in a mathematical interpretation, $G\left(x,y|x',y'\right)$ is nothing but the kernel which plays the role of transformation function between the probability densities $w_{\eta} \left(x,y\right)\; \, ,\quad \eta =\left\{in,\, out\right\}$, from initial version ($\eta =in$) into  final reading ($\eta =out$). Due to the respective interpretation as well as to the evident normalization conditions
\begin{equation} \label{eq:5)} 
\int _{\Omega _{X} }\int _{\Omega _{Y} }w_{\eta } \left(x,y\right)  \cdot dx\cdot dy=1 
\end{equation} 
The kernel $G\left(x,y|x',y'\right)$ must satisfy the following relations    
\begin{equation} \label{eq:6)} 
\begin{split} \int _{\Omega _{X} }\int _{\Omega _{Y} }G\left(x,y|x',y'\right)\cdot \,   dx\cdot dy &=1
\\ 
\int _{\Omega _{X'}} \int _{\Omega _{Y'}} G\left(x,y|x',y'\right)\cdot \,   dx'\cdot dy' &=1 
\end{split} 
\end{equation} 
where, according to the above stipulations, we take 
$\Omega _{X} \equiv \Omega _{X'} \; ;\; \Omega _{Y} \equiv \Omega _{Y'}$.

Now we can estimate the following examples of mean (expected) values 
\begin{equation} \label{eq:7} 
\begin{split} \left\langle X\right\rangle _{\eta } &=\int _{\Omega _{X} }\int _{\Omega _{Y} }x\cdot w_{\eta } \left(x,y\right)  \cdot dx\cdot dy \\ 
\left\langle A\right\rangle _{\eta } =\left\langle f\left(X,Y\right)\right\rangle _{\eta } &=\int _{\Omega _{X} }\int _{\Omega _{Y} }f\left(x,y\right)\cdot w_{\eta } \left(x,y\right)  \cdot dx\cdot dy \end{split} 
\end{equation} 
and similar expressions for the mean values $\left\langle Y\right\rangle _{\eta } $ and $\left\langle B\right\rangle _{\eta }$ of the observables $\mathcal{Y}$ respectively $\mathcal{B}$.

As regards the second relation from \eqref{eq:7} it is the place here to specify the fact that, mathematically, in general $\left\langle f\left(X,Y\right)\right\rangle _{\eta }$ is not equal to 
$f\left(\left\langle X\right\rangle _{\eta } ,\left\langle Y\right\rangle _{\eta } \right)$ 
(see \cite{18}).

By taking into account the above introduced notations one can see that, in the spirit of relations \eqref{eq:7}, the symbols $\left\langle \mathcal{Z}_{\alpha } \right\rangle _{\eta }$ will stand for the $\eta $-versions of the mean (expected) values of the observables $\mathcal{Z}_{\alpha}$. 
Then,  for the respective observables,  the measuring errors (``uncertainties''), induced by the discussed kind of measurement, can be evaluated theoretically by the following \textit{first order indicators}:
\begin{equation} \label{eq:8}  
\mathcal{PEI}  \left\{ \left\langle Z_{\alpha } \right\rangle \right\} = \left\langle Z_{\alpha } \right\rangle _{out} -\left\langle Z_{\alpha } \right\rangle _{in} 
\end{equation} 
where the symbol $\mathcal{PEI} \left\{ Q \right\}$ signifies the \textit{predicted error indicator} of the quantity $Q$.

Note that, above and in the following discussions, in respect with the indicators of measuring errors, we adopted the adjective \textit{``predicted''} (or \textit{``theoretically estimared''}) because all  of our considerations consist in a theoretical (mathematical) modelling of a measuring process. Or within such a modelling we are dealing only with theoretical (mathematical) elements presumed to reflect in a plausible manner all the main characteristics of the considered process. On the other hand, comparatively, in experimental physics for the indicators regarding the measuring errors it is recommendatory to use the adjective \textit{``factual''}, i.e the notation  
$\mathcal {FEI} \left\{ Q \right\}$ for the \textit{factual error indicator} of the quantity $Q$. 
This because the respective kind of errors are obtained from factual lots of experimental data.

Now is the place to specify that, because the observables $\mathcal{Z}_{\alpha}$ are variables of random kind, a more complete description of their measuring errors (``uncertainties'') must be done not only by the first order indicators defined in \eqref{eq:8}. The alluded description requires to resort also to the class of \textit{superior indicators} (expressible in terms of probabilistic higher order centered moments). From the mentioned class, for the here discussed observables, we  will focus only on the errors regarding the following  second order probabilistic moments:
\textit{variances} $\mathrm{Var}_{\eta } \left(Z_{\alpha } \right)$ (or, equivalently, the 
\textit{standard deviations} $\sigma _{\eta } \left(Z_{\alpha } \right)$), respectively the 
\textit{covariances} $\mathrm{Cov}_{\eta } \left(Z_{\alpha }, Z_{\beta } \right)\, , \; 
\left(\alpha \ne \beta \right)$. The mentioned moments are defined thruogh the relations
\begin{equation} \label{eq:9} 
\mathrm{Var}_{\eta } \left(Z_{\alpha } \right)=\sigma _{\eta } ^{2} \left(Z_{\alpha } \right)=\left\langle \left(\mathcal{Z}_{\alpha } -\left\langle Z_{\alpha } \right\rangle _{\eta } \right)^{2} \right\rangle _{\eta }  
\end{equation} 
\begin{equation} \label{eq:10} 
\mathrm{Cov}_{\eta } \left(Z_{\alpha } ,Z_{\beta } \right)=\left\langle \left(Z_{\alpha } -\left\langle
Z_{\alpha } \right\rangle _{\eta } \right)\cdot \left(Z_{\beta } -\left\langle Z_{\beta } \right\rangle _{\eta } \right)\right\rangle _{\eta } \, ,\quad \alpha \ne \beta  
\end{equation} 
where the mean (expected) values $\left\langle \cdots \right\rangle _{\eta }$ have the significances given in \textit{\eqref{eq:7}}.

For the moments defined in \eqref{eq:9} and \eqref{eq:10} the predicted errors can be appreciated by means of the following indicators
\begin{equation} \label{eq:11} 
\mathcal{PEI} \left\{\mathrm{Var}\left(Z_{\alpha } \right)\right\}
=\mathrm{Var}_{out} \left(Z_{\alpha } \right)-\mathrm{Var}_{in} \left(Z_{\alpha } \right) 
\end{equation} 
\begin{equation} \label{eq:12} 
\mathcal{PEI}  \left\{\sigma \left(Z_{\alpha } \right)\right\}=\sigma _{out} \left(Z_{\alpha } \right)-\sigma _{in} \left(Z_{\alpha } \right) 
\end{equation} 
\begin{equation} \label{eq:13} 
\mathcal{PEI} \left\{\mathrm{Cov}\left(Z_{\alpha } ,Z_{\beta } \right)\right\}
=\mathrm{Cov}_{out} \left(Z_{\alpha } ,Z_{\beta } \right)-\mathrm{Cov}_{in} \left(Z_{\alpha },
Z_{\beta } \right)\, ,\quad \alpha \ne \beta  
\end{equation} 
Note that according to some concrete examples (see below), for variance and standard deviation, the error indicators $\mathcal{PEI} \left\{\mathrm{Var}\left(Z_{\alpha } \right)\right\}$ respectively
$\mathcal{PEI}$ $\left\{\sigma \left(Z_{\alpha } \right)\right\}$ seem to be positive quantities.

In the above considerations about the subject of measuring error (uncertainties) we regarded only the second order probabilistic moments (variances and covariances). It is the place here to mention that recently \cite{19}, in connection with the same subject, was proposed to use also probabilistic moments of higher order (3,4,5 and even 6). Speaking of the respective proposal we add here the fact that our considerations can be extended without any problem by referring to the higher order probabilistic moments. So we can resort to the \textit{higher s-order covariances}
\begin{equation} \label{eq:14} 
\mathrm{Cov}_{\eta } \left(Z_{\alpha}^{m}, Z_{\beta } ^{n} \right)=\left\langle \left(Z_{\alpha } -\left\langle Z_{\alpha } \right\rangle _{\eta } \right)^{m} \cdot \left(Z_{\beta } -\left\langle 
Z_{\beta } \right\rangle _{\eta } \right)^{n} \right\rangle _{\eta } \, ,\quad \left\{\begin{array}{l} {\alpha = \text{or} \ne \beta } \\ {m+n=s\ge 3} \end{array}\right.  
\end{equation} 
respectively to their \textit{predicted error indicators}
\begin{eqnarray} \label{eq:15} 
\mathcal{PEI} \, \left\{\mathrm{Cov}\left(Z_{\alpha } ^{m} ,Z_{\beta } ^{n} \right)\right\}=
\mathrm{Cov}_{out} \left(Z_{\alpha } ^{m} ,Z_{\beta } ^{n} \right)-\mathrm{Cov}_{in} \left(Z_{\alpha } ^{m} ,Z_{\beta } ^{n} \right) \nonumber \\ 
\left\{\begin{array}{l} {\alpha = \text{or} \ne \beta } \\ {m+n=s\ge 3} \end{array}\right.
\end{eqnarray}   
Now is the place to specify that the predicted indicators defined through the relations   
\eqref{eq:11}-\eqref{eq:13} and \eqref{eq:15} with fixed first \textit{s} orders (i.e. with a finite and not very high value for $s$) give in fact a somehow \textit{truncated description} of the measuring errors (uncertainties). A somehow \textit{more  comprehensive description} of the  error indicators can be done by extending one of our old idea \cite{6}, \cite{9}, through the following \textit{Shanon's information entropies}
\begin{equation} \label{eq:16)} 
\mathcal H_{\eta } =-\int _{\Omega _{X} }dx \int _{\Omega _{Y} }dy\cdot  w_{\eta } \left(x,y\right)\cdot \ln w_{\eta } \left(x,y\right) 
\end{equation} 
Note  that in these expressions for the entropies $\mathcal H_{\eta }$ the variables $x$ and 
$y$ must be considered only by their \textit{dimensionless values} (i.e. without the corresponding physical dimensions of their units).

Then as a more comprehensive description  regarding  the measuring  errors can be done through  the following predicted indicator
\begin{equation} \label{eq:17} 
\mathcal{PEI} \left\{\mathcal H\right\}={\rm {\mathcal H}}_{out} -{\rm {\mathcal H}}_{in}  
\end{equation} 
By using some simple calculations, completely similar with the ones given in \cite{6,9},
for one-observable situation, one finds easily
\begin{equation} \label{eq:18} 
\mathcal{PEI} \left\{{\rm {\mathcal H}}\right\}\ge 0 
\end{equation} 
The cases with $\mathcal{PEI} \left\{\mathcal H\right\}=0$ correspond to the \textit{ideal measurement} when, as we noted above, the kernel $G\left(x,y|x',y'\right)$ is of the particular form $G\left(x,y|x',y'\right)=\delta \left(x-x'\right)\cdot \delta \left(y-y'\right)$. On the other hand  the cases in which  $\mathcal{PEI} \left\{{\rm {\mathcal H}}\right\}>0$ are referring to the cases of \textit{non-ideal measurements} for which the kernel $G\left(x,y|x',y'\right)$ does not have the above noted particular form.

The above presented general considerations are illustrated through a very simple situation done bellow in the Appendix A.

\section{The case of quantum  observables}

As it is known random characteristics are manifested also by the quantum observables (encountered in quantum mechanics \cite{20} and in quantum statistics \cite{5}). In this section we intend to promote discussions about the description of measurements regarding such observables. We will develop our intention by implementation of some ideas from the previous section connected with the classical random observables.

In order to be more explicit we consider a simple microparticle (quantum system) endowed only with orbital motions (characteristics). The state of such a system is described by the wave function 
$\Psi \left(\vec{r},t\right)$ having the following probabilistic direct and subsequent significances. 

The quantity 
\begin{equation} \label{eq:19} 
dP=\left|\Psi \left(\vec{r},t\right)\right|^{2} d^{3} \vec{r}=\rho \left(\vec{r},t\right)d^{3} \vec{r} 
\end{equation} 
denotes the probability that, at the moment $t$, the particle to be present in the infinitesimal volume $d^{3} \vec{r}$ in the neighborhood of the point $\vec{r}$. Consequently the quantity
$\rho \left(\vec{r},t\right)$ plays the role of a \textit{probability density}. But
$\rho \left(\vec{r},t\right)$ describes only the presence of the particle in a location in space but not the \textit{travel} of the same particle through the respective location. Such a travel is described by the probability current defined as follows
\begin{equation} \label{eq:20} 
\vec{j}\left(\vec{r},t\right)=-\frac{i\hbar}{2m_{0}} \left[\Psi^{*} \left(\vec{r},t\right)\cdot \nabla \Psi \left(\vec{r},t\right)-\Psi \left(\vec{r},t\right)\cdot \nabla \Psi^{*} \left(\vec{r},t\right)\right] 
\end{equation} 
where $m_0$ denotes the mass of the microparticle.

The  physical properties of the particle are associated with the observables $A_{\alpha} \, ,\; \left(\alpha =1,2,\cdots ,n\right)$, described by the quantum operators $\hat{A}_{\alpha}$. The respective operators are generalized random variables in the sense that for the particle in a given state 
(described by $\Psi \left(\vec{r},t\right)$) they are characterized generally by a spectrum (a number) of values.

By reporting our discussions to the above reminded quantum notions now we try to develop a description of quantum measurements by adopting some viewpoints presented in the previous section about classical measurements. Firstly we note that a measurement of a random observable does not change its spectrum of values. This fact means that a quantum measurement must be regarded as an action that preserves the mathematical expressions of the operators $\hat{A}_{\alpha}$. On the other hand, by taking into account the imperfections of experimental devices, it is credible and rightful the idea that the same action must be regarded also as a process in which the \textit{input (in)} information (probabilities) regarding the intrinsic properties of the measured system are converted in \textit{output (out)} information incorporated within the data received on a recorder device. So a quantum measurement appears also as an \textit{information transmission process}. The respective process must be depicted as a modification of the form
\begin{equation} \label{eq:21} 
\Psi _{in} \left(\vec{r},t\right)\to \Psi _{out} \left(\vec{r},t\right) 
\end{equation} 
for the wave function $\Psi \left(\vec{r},t\right)$. By taking into account the relations 
\eqref{eq:19} and \eqref{eq:20} it is possible that the modification \eqref{eq:21} to be described by means of changes for both the probability quantities $\rho \left(\vec{r},t\right)$ and $\vec{j}\left(\vec{r},t\right)$. Because, as a rule the devices which measures the quantities $\rho \left(\vec{r},t\right)$ and $\vec{j}\left(\vec{r},t\right)$ are technically distinct objects, the mentioned changes must be described through separate transformations of the forms
\begin{equation} \label{eq:22} 
\rho _{in} \left(\vec{r},t\right)\to \rho _{out} \left(\vec{r},t\right) 
\end{equation} 
\begin{equation} \label{eq:23} 
\vec{j}_{in} \left(\vec{r},t\right)\to \vec{j}_{out} \left(\vec{r},t\right) 
\end{equation} 
\textit{Observation}: Note that, because in fact $\vec{j}_{\eta } \left(\vec{r},t\right)\, ,\; (\eta =in,\; out)$ are vectors, the relation \eqref{eq:23} consists in a set of inter-connected transformations among the Cartesian components $j_{\eta ;\mu } \left(\vec{r},t\right)\, ,\; (\mu =x,y,z)$ of the respective vectors (see below the formulas \eqref{eq:25}).

In the above relations the quantities $\rho _{in} \left(\vec{r},t\right)$ and  
$\vec{j}_{in} \left(\vec{r},t\right)$ describes the intrinsic properties of the measured system. 
On the other hand the quantities $\rho _{out} \left(\vec{r},t\right)$ and  $\vec{j}_{out} \left(\vec{r},t\right)$  incorporate the information about the same system, but obtained on the  measuring recorder device.

Similarly with the classical situations for the description of the quantum measurements we have to take into account the fact that, mainly, the  measuring devices must have the same properties \textbf{\textit{P1}} and \textbf{\textit{P2}} mentioned in  previous section.

Based on the above noted facts, as well as the \textit{observation} associated with  
\eqref{eq:23}, we consider that the transformations \eqref{eq:22} and  \eqref{eq:23} 
can be taken of the forms
\begin{equation} \label{eq:24} 
\rho_{out} \left(\vec{r},t\right)=\int_{\mathbb{R}^{3}}\Gamma \left(\vec{r}\, | \vec{r'}\right) \cdot \rho _{in} \left(\vec{r'},t\right)d^{3} \vec{r'} 
\end{equation} 
\begin{equation} \label{eq:25} 
j_{out;\mu} \left(\vec{r},t\right)=\sum _{\nu =1}^{3}\int _{\mathbb{R}^{3}}\Lambda _{\mu \nu } \left(\vec{r}\,|\vec{r'}\right) \cdot j_{in;\nu } \left(\vec{r'},t\right)d^{3} \vec{r'} \, ,\quad \mu ,\nu =x,y,z 
\end{equation} 
Note that by omitting the time $t$ in the expressions of kernels $\Gamma \left(\vec{r}\,|\vec{r'}\right)$ and $\Lambda _{\mu \nu } \left(\vec{r}\,|\vec{r'}\right)$, in our model, we consider  the measurements as instantaneous actions (i.e. we neglect the relativistic effects connected with retarded influences).

In the relations \eqref{eq:24} and \eqref{eq:25}, consonantly with the viewpoint of physics, the kernels $\Gamma \left(\vec{r}\,|\vec{r'}\right)$ and $\Lambda _{\mu \nu } \left(\vec{r}\,|\vec{r'}\right)$ depict theoretically the actions of the measuring devices. For ideal devices able to assure 
$\rho _{out} \left(\vec{r},t\right)\equiv \rho _{in} \left(\vec{r},t\right)$ and $j_{out;\mu } \left(\vec{r},t\right)\equiv j_{in;\mu } \left(\vec{r},t\right)$ the mentioned kernels must have the forms 
$\Gamma \left(\vec{r}\,|\vec{r'}\right)=\delta \left(\vec{r}-\vec{r'}\right)$ respectively $\Lambda _{\mu \nu } \left(\vec{r}\,|\vec{r'}\right)=\delta \left(\vec{r}-\vec{r'}\right)\cdot \delta _{\mu \nu } $  (where $\delta (\vec{\xi })$  denotes the 3D Dirac's  $\delta$-function of the vectorial argument $\vec{\xi }$ and $\delta _{\mu \nu }$ signifies the Kronecker delta).

On the other hand, in a mathematical regard, the kernels $\Gamma \left(\vec{r}\,|\vec{r'}\right)$ and 
$\Lambda _{\mu \nu } \left(\vec{r}\,|\vec{r'}\right)$  play the role of  transformation functions among  the  probabilistic densities $\rho _{\eta } \left(\vec{r},t\right)$ respectively currents 
$j_{\eta ;\mu } \left(\vec{r},t\right)$, from initial versions ($\eta=in$) into  final  readings 
($\eta=out$). Due to  the respective regard  as well as to the naturally  implied probabilistic normalizations the kernels $\Gamma \left(\vec{r}\,|\vec{r'}\right)$ and $\Lambda _{\mu \nu } \left(\vec{r}\,|\vec{r'}\right)$ must satisfy the following conditions:
\begin{equation} \label{eq:26} 
\int _{\mathbb{R}^{3}}\Gamma \left(\vec{r}\,|\vec{r'}\right) \cdot d^{3} \vec{r}=\int_{\mathbb{R}^{3}}\Gamma \left(\vec{r}\,|\vec{r'}\right) \cdot d^{3} \vec{r'}=1 
\end{equation} 
\begin{equation} \label{eq:27} 
\sum _{\mu =1}^{3}\int _{\mathbb{R}^{3}}\Lambda_{\mu \nu } \left(\vec{r}\,|\vec{r'}\right) \cdot d^{3} \vec{r} =\sum _{\nu =1}^{3}\int _{\mathbb{R}^{3}}\Lambda _{\mu \nu } \left(\vec{r}\,|\vec{r'}\right) \cdot d^{3} \vec{r'} =1 
\end{equation} 
Associated to the two kinds of situations $\eta =in,\, out$, for a set $A_{\alpha }\, , \; (\alpha =1,2,\cdots ,n)$ of quantum observables described by the operators $\hat{A}_{\alpha}$, the following probabilistic parameters can be evaluated
\begin{equation} \label{eq:28} 
<A_{\alpha}>_{\eta } = \int_{\mathbb{R}^{3}} \Psi_{\eta }^{*} \left(\vec{r},t\right)\hat{A}_{\alpha } \Psi_{\eta } \left(\vec{r},t\right) \cdot d^{3} \vec{r} 
\end{equation} 
\begin{equation} \label{eq:29} 
\mathrm{Var}_{\eta } \left( A_{\alpha } \right) =\sigma _{\eta } ^{2} \left( A_{\alpha } \right)
= \left\langle \left( A_{\alpha} - <A_{\alpha}>_{\eta } \right)^{2} \right\rangle
\end{equation} 
\begin{equation} \label{eq:30} 
\mathrm{Cov}_{\eta } \left( A_{\alpha} , A_{\beta} \right) =\left\langle\left( A_{\alpha } -<A_{\alpha} >_{\eta } \right)\left(A_{\beta } -<A_{\beta }>_{\eta } \right)\right\rangle_{\eta } \, , \quad 
\alpha \ne \beta  
\end{equation} 
\begin{equation} \label{eq:31} 
\mathrm{Cov}_{\eta } \left(A_{\alpha }^{m}, A_{\beta }^{n} \right)=\left\langle 
\left(A_{\alpha } - <A_{\alpha}>_{\eta} \right)^{m} \left(A_{\beta } -<A_{\beta}>_{\eta } \right)^{n} 
\right\rangle \; ,\quad \left\{\begin{array}{l} {\alpha =\text{or} \ne \beta} \\ {m+n=s\ge 3} 
\end{array}\right.  
\end{equation} 
Here we have to elucidate the folloving apparently intriguing fact. On the one hand, the evaluation of the parameters \eqref{eq:28}-\eqref{eq:31} requires the knowledge of the wave functions $\Psi _{\eta } \left(\vec{r},t\right)$. On the other hand in our above considerations about quantum measurements we use the probability densities and currents $\rho _{\eta } \left(\vec{r},t\right)$ respectively  
$\vec{j}_{\eta } \left(\vec{r},t\right)$. The elucidation can be done as follows.

If the operators $\hat{A}_{\alpha}$ do not depend on $\nabla$ (i.e. $\hat{A}_{\alpha } =A_{\alpha } \left(\vec{r}\right)$) in evaluating the integrals from \eqref{eq:28} one can use the evident equality
\begin{equation} \label{eq:32} 
\Psi _{\eta }^{*} \left(\vec{r},t\right)\hat{A}_{\alpha} \Psi _{\eta } \left(\vec{r},t\right)=
A_{\alpha} \left(\vec{r}\right)\cdot \rho_{\eta } \left(\vec{r},t\right) 
\end{equation} 
When the operators $\hat{A}_{\alpha } $ depend on nabla-operator $\nabla $ (i.e. $\hat{A}_{\alpha } =A_{\alpha } \left(\nabla \right)$) we can resort to one of the next relations
\begin{equation} \label{eq:33} 
\Psi _{\eta }^{*} \left(\vec{r},t\right)\nabla \Psi _{\eta } \left(\vec{r},t\right)=\frac{1}{2} \nabla \rho _{\eta } \left(\vec{r},t\right)+\frac{im_{0} }{\hbar } \vec{j}_{\eta } \left(\vec{r},t\right) 
\end{equation} 
\begin{eqnarray} \label{eq:34} 
\Psi _{\eta }^{*} \left(\vec{r},t\right)\nabla^{2} \Psi _{\eta } \left(\vec{r},t\right)
&=&\sqrt{\rho _{\eta } \left(\vec{r},t\right)} \cdot \nabla ^{2} \sqrt{\rho _{\eta } \left(\vec{r},t\right)} + \nonumber \\ 
&&+\frac{i m_{0}}{\hbar } \nabla \, \vec{j}_{\eta } \left(\vec{r},t\right)-\frac{m_{0} ^{2} }{\hbar^2} \cdot \frac{ \vec{j}_{\eta }^{2} \left(\vec{r},t\right)}{\rho _{\eta } ^{2} \left(\vec{r},t\right)}  
\end{eqnarray}
Now, within the above promoted model about the description of measurements, let us note what are the indicators of the \textit{predicted errors} ($\mathcal{PEI}$), specific for quantum measurements. Taking into account the formulas \eqref{eq:28}-\eqref{eq:31} now we can state that the roles of such indicators can be performed by the following quantities
\begin{equation} \label{eq:35} 
\mathcal{PEI} \left\{ < A_{\alpha } > \right\} = <A_{\alpha }>_{out} - <A_{\alpha}>_{in}  
\end{equation} 
\begin{equation} \label{eq:36} 
\mathcal{PEI} \left\{ \mathrm{Var} \left( A_{\alpha } \right) \right\}=
\mathrm{Var}_{out} \left(A_{\alpha } \right)- \mathrm{Var}_{in} \left(A_{\alpha } \right) 
\end{equation} 
\begin{equation} \label{eq:37} 
\mathcal{PEI} \left\{ \sigma \left(A_{\alpha } \right)\right\}
=\sigma_{out} \left(A_{\alpha } \right)-\sigma _{in} \left(A_{\alpha } \right) 
\end{equation} 
\begin{equation} \label{eq:38} 
\mathcal{PEI} \left\{\mathrm{Cov}\left(A_{\alpha } ,A_{\beta } \right)\right\}=\mathrm{Cov}_{out} 
\left(A_{\alpha } ,A_{\beta } \right)-\mathrm{Cov}_{in} \left(A_{\alpha } ,A_{\beta } \right)\, ,
\quad \alpha \ne \beta  
\end{equation} 
\begin{eqnarray} \label{eq:39}
\mathcal{PEI} \left\{\mathrm{Cov}_{\eta } \left(A_{\alpha } ^{m} ,A_{\beta } ^{n} \right)\right\}=
\mathrm{Cov}_{out} \left(A_{\alpha } ^{m} ,A_{\beta } ^{n} \right)-\mathrm{Cov}_{in} \left(A_{\alpha } ^{m} ,A_{\beta } ^{n} \right)\, , \nonumber \\ 
\left\{\begin{array}{l} {\alpha = \text{or} \ne \beta } \\ {m+n=s\ge 3} \end{array}\right.  
\end{eqnarray} 
In the end of this section we wish to note that, similarly with the case of classical measurements (discussed in Section 2), the measuring errors specific for quantum measurements can be described through the entropies
\begin{equation} \label{eq:40} 
\mathcal H \left(\rho _{\eta } \right)=-\int _{\mathbb{R}^{3}}\rho _{\eta } \left(\vec{r},t\right) \cdot \ln \rho _{\eta } \left(\vec{r},t\right)\cdot d^{3} \vec{r} 
\end{equation} 
\begin{equation} \label{eq:41} 
\mathcal H \left(\vec{j}_{\eta } \right)=-\int _{\mathbb{R}^{3}}\left|\vec{j}_{\eta } \left(\vec{r},t\right)\right| \cdot \ln \left|\vec{j}_{\eta } \left(\vec{r},t\right)\right|\cdot d^{3} \vec{r} 
\end{equation} 
Add here the observation  that in the above  expressions for the entropies $\mathcal H \left(\rho _{\eta } \right)$ and $\mathcal H \left(\vec{j}_{\eta } \right)$ the variables $\rho _{\eta } \, ,\; \vec{j}_{\eta } \, ,\; d^{3} \vec{r}$ must be considered only by their dimensionless values (i.e. without the corresponding physical dimensions of their units).

Now we note the fact that for the entropies $\mathcal H \left(\rho _{\eta } \right)$ and $\mathcal H\left(\vec{j}_{\eta } \right)$ their \textit{predicted errors} ($\mathcal{PEI}$) are described by the indicators defined through the following relations
\begin{equation} \label{eq:42} 
\mathcal{PEI} \left\{ \mathcal H \left(\rho \right)\right\}=\mathcal H\left(\rho _{out} \right)-\mathcal H \left(\rho _{in} \right) 
\end{equation} 
\begin{equation} \label{eq:43} 
\mathcal{PEI} \left\{ \mathcal H \left(\vec{j}\right)\right\}= \mathcal H \left(\vec{j}_{out} \right) - \mathcal H \left(\vec{j}_{in} \right) 
\end{equation} 
In the end of this section it must be specified the fact that all the indicators \eqref{eq:35}-\eqref{eq:39}, \eqref{eq:42} and \eqref{eq:43} are null quantities in the cases of ideal  measurements (i.e. when, as it was noted above, in the measurement description we operate with the kernels $\Gamma \left(\vec{r}\, |\vec{r'}\right)=\delta \left(\vec{r}-\vec{r'}\right)$ respectively $\Lambda _{\mu \nu } \left(\vec{r}\,|\vec{r'}\right)=\delta \left(\vec{r}-\vec{r'}\right)\cdot \delta _{\mu \nu }$).  
The same indicators are non-null quantities in the cases of non-ideal measurements.

\section{Conclusions}

In this paper  we have suggested that the measurements regarding the cases of classical respectively quantum observables to be described theoretically in somewhat similar manners.

Our suggestion  is builded on the idea that, by means of a simple settings, in both cases the observables can be described mathematically as random variables. In such a description a measurement preserves the values spectrum of an observable but changes the corresponding probabilistic weights (classical probability density or, respectively, the quantum wave function). 

The mentioned change is modeled in the our previous two sections as follows. An \textit{input (in)} information about the intrinsic properties of the measured system is converted into an \textit{output (out)} information incorporated in the data received on a recorder device. So a measurement can be regarded as a particular example of an \textit{information transmission process}.

Our  modeling of measurements ensures theoretical estimations for \textit{predicted error indicators} specific to the mean values as well as to the fluctuations of both classical and quantum  types of observables. 

The measuring modeling approached here stands out of ourdays prevalent opinion that measurement theories must be depicted in entirely different manners in the classical and quantum cases. Particularly the respective modeling does not have any natural motivation to refer to an \textit{indubitable connection} between the quantum measurements and the (Heisenberg's) uncertainty relations. For more  details and arguments regarding the alluded connection as well as the whole class of  questions about the interpretation of the uncertainty relations see the works \cite{7}, \cite{8} and \cite{21} of one of us (S.D.).

\appendix

\section{A simple illustration for a single  classical randon observable.}

For illustrating the general considerations from the above Section 2 now let us present in its details the following simple example. We regard the situation of a measured system, such are the ones studied in the phenomenological theory of fluctuations \cite{4}. We take the system as described by a single random observable $\mathcal{X}$. In respect with the observable $\mathcal{X}$ the inner state of the system is decribed by the Gaussian type \textit{in}-probability density.
\begin{equation} \label{eq:44} 
w_{in} \left(x\right)\propto \exp \left\{-\frac{x^{2} }{2\sigma _{in} ^{2} } \right\} 
\end{equation} 
Note that in this Appendix, as well as in the next one, for simplicity of writings, we omit the explicit mention of the normalizing constants for the probabilistic entities. Such constants can be easily introduced by the interested readers. 

We consider also the case of a measuring device endowed with a good accuracy. Then, accordingly with the usual ideas about measurements, the corresponding kernel $G\left(x|x'\right)$ can be taken of the Gaussian form
\begin{equation} \label{eq:45} 
G\left(x|x'\right)\propto \exp \left\{-\frac{\left(x-x'\right)^{2} }{2\sigma _{D} ^{2} } \right\} 
\end{equation} 
where $\sigma_{D}$ denotes the \textit{precision indicator} of the measuring device. 
With the expressions \eqref{eq:44} and \eqref{eq:45} by using a transformation of type 
\eqref{eq:4} one finds
\begin{equation} \label{eq:46} 
w_{out} \left(x\right)\propto \exp \left\{-\frac{x^{2} }{2\sigma _{out} ^{2} } \right\} \, ,\quad \sigma _{out} ^{2} =\sigma _{in} ^{2} +\sigma _{D} ^{2}  
\end{equation} 
Then, for the considered measurement, the predicted errors regarding the fluctuations of the random observable $\mathcal{X}$ are described by the following indicators  
\begin{equation} \label{eq:47} 
\mathcal{PEI}  \left\{\sigma \right\}=\sqrt{\sigma _{in} ^{2} +\sigma _{D} ^{2} } -\sigma _{in}  
\end{equation} 
\begin{equation} \label{eq:48} 
\mathcal{PEI} \left\{ \mathcal{H} \right\}=\ln \frac{\sqrt{\sigma _{in} ^{2} +\sigma _{D} ^{2} } }{\sigma _{in} }  
\end{equation} 
In the case when the measuring device is of very high precision $\sigma _{D} \ll \sigma _{in}$ and the above indicators become
\begin{equation} \label{eq:49} 
\mathcal{PEI} \left\{\sigma \right\}\approx \frac{1}{2} \left(\frac{\sigma _{D} }{\sigma _{in} } \right)\, , \quad \mathcal{PEI} \left\{ \mathcal{H} \right\}\approx 2\left(\frac{\sigma _{D} }{\sigma _{in} } \right) 
\end{equation} 
For the situation when $\sigma _{D} \to 0$ one observes that $G\left(x|x'\right)\to \delta \left(x-x'\right)$ and $\mathcal{PEI} \left\{\sigma \right\}\to 0$ respectively $\mathcal{PEI} \left\{ \mathcal H\right\}\to 0$. Evidently that for fluctuations of the random observable $\mathcal{X}$, such a situation corresponds to an ideal measurement i.e. without any error (uncertainty).

\section{A simple illustration regarding the quantum observables}

For a simple exemplification of the model presented in Section 3 let us refer to a microparticle in a one-dimensional motion along the x-axis. We take 
\begin{equation} \label{eq:50} 
\Psi _{in} \left(\vec{r},t\right)\equiv \psi _{in} \left(x\right)=\left|\psi _{in} \left(x\right)\right|\cdot \exp \left\{i\phi _{in} \left(x\right)\right\} 
\end{equation} 
with
\begin{equation} \label{eq:51} 
\left|\psi _{in} \left(x\right)\right|\propto \exp \left\{-\frac{\left(x-x_{0} \right)^{2} }{4\sigma _{in} ^{2} } \right\}\, ,\quad \phi _{in} \left(x\right)=kx 
\end{equation} 
In this Appendix, in order to simplify the formulas, we omit also the explicit writing of the normalizing constants for the probabilistic entities. Such constants can be easily introduced by the interested readers.

Correspondingly to the relations \eqref{eq:50} and \eqref{eq:51} we have the expressions
\begin{equation} \label{eq:52} 
\rho _{in} \left(x\right)=\left|\psi _{in} \left(x\right)\right|^{2} \, ,\quad j_{in} \left(x\right)=\frac{\hbar k}{m_{0} } \left|\psi _{in} \left(x\right)\right|^{2}  
\end{equation} 
These expressions show that the the intrinsic properties of the measured  microparticle are described by the parameters $x_0$, $\sigma _{in}$ and $k$. If the errors induced by measuring processes are small the transformation kernels $\Gamma$ and $\Lambda$ in \eqref{eq:26} and \eqref{eq:27} can be considered of Gaussian forms like
\begin{equation} \label{eq:53} 
\Gamma \left(x|x'\right)\propto \exp \left\{-\frac{\left(x-x'\right)^{2} }{2\gamma ^{2} } \right\} 
\end{equation} 
\begin{equation} \label{eq:54} 
\Lambda \left(x|x'\right)\propto \exp \left\{-\frac{\left(x-x'\right)^{2} }{2\lambda ^{2} } \right\} \end{equation} 
where $\gamma$ and $\lambda$ describe the characteristics of the measuring devices. Then for
$\rho _{out}$ and $j_{out}$ one finds
\begin{equation} \label{eq:55} 
\rho _{out} \left(x\right)\propto \exp \left\{-\frac{\left(x-x_{0} \right)^{2} }{2\left(\sigma _{in} ^{2} +\gamma ^{2} \right)} \right\} 
\end{equation} 
\begin{equation} \label{eq:56} 
j_{out} \left(x\right)\propto \hbar k\cdot \exp \left\{-\frac{\left(x-x_{0} \right)^{2} }{2\left(\sigma _{in} ^{2} +\gamma ^{2} \right)} \right\} 
\end{equation} 
It can be seen that in the case when both $\gamma \to 0$ and $\lambda \to 0$ the kernels $\Gamma \left(x|x'\right)$ and $\Lambda \left(x|x'\right)$ degenerate into the Dirac's function $\delta(x-x')$. Then $\rho _{in} \to \rho _{out}$ and $j_{in} \to j_{out}$. Such a case corresponds to ideal measurements. Alternatively the  cases when $\gamma \ne 0$ and/or $\lambda \ne 0$ are associated with non-ideal measurements.

As observables of interest we take $A_{1} =x =$ the first of Cartesian  coordinates and  $A_{2} =p =$
 the first of Cartesian momenta. The respective observables are described by the operators $\hat{x}=x\cdot$ and $\hat{p}=-i\hbar \frac{\partial }{\partial x}$. Then, according to the scheme presented in Section 3 and through some simple calculations, one can evaluate the indicators specific for the \textit{predicted errors} ($\mathcal{PEI}$) regarding the considered measurements of the mentioned observables. The expressions of the respective indicators are given by the relations
\begin{equation} \label{eq:57} 
\mathcal{PEI} \left\{ <x> \right\}=0\, ,\quad \mathcal{PEI} \left\{ <p> \right\}=0\, ,\quad 
\mathcal{PEI} \left\{ \mathrm{Cov} \left(x,p\right)\right\}=0 
\end{equation} 
\begin{equation} \label{eq:58} 
\mathcal{PEI} \left\{ \mathrm{Var} \left(x\right)\right\}=\gamma ^{2}  
\end{equation} 
\begin{equation} \label{eq:59}
\mathcal{PEI} \left\{ \mathrm{Var} \left(p\right)\right\}=\frac{\hbar ^{2} k^{2} }{\left(\sigma _{in} ^{2} +\lambda ^{2} \right)\left(\sigma _{in} ^{2} +2\gamma ^{2} -\lambda ^{2} \right)} -\frac{\hbar ^{2} k^{2} }{\left(\sigma _{in} ^{2} +\gamma ^{2} \right)} -\left(\sigma _{in} ^{2} +\gamma ^{2} \right) \end{equation} 

In the same modeling of measurements the predicted errors regarding the informational entropies are characterized by the indicators
\begin{equation} \label{eq:60} 
\mathcal{PEI} \left\{ \mathcal{H} \left(\rho \right)\right\}=\frac{1}{2} \ln \left(1+\frac{\gamma ^{2} }{\sigma _{in} ^{2} } \right) 
\end{equation} 
\begin{equation} \label{eq:61} 
\mathcal{PEI} \left\{ \mathcal{H} \left(j\right)\right\}=\frac{1}{2} \ln \left(1+\frac{\lambda ^{2} }
{\sigma _{in} ^{2} } \right) 
\end{equation} 

Now, in our illustration started with the exemplified through the relations \eqref{eq:50} and 
\eqref{eq:51}, let us restrict to the situation when $x_{0} =0 ,\, k=0$ and $\sigma _{in} =\sqrt{\frac{\hbar }{2m_{0} \omega } }$. Then the considered system is just a linear oscillator in its ground state ($m_0$ = mass and $\omega$ = angular frequency). As observable of interest we consider the energy of the oscillator described by the Hamiltonian operator
\begin{equation} \label{eq:62} 
\hat{H}=-\frac{\hbar ^{2} }{2m_{0} } \frac{d^{2} }{dx^{2} } +\frac{m_{0} \omega ^{2} }{2} x^{2}  
\end{equation} 
It is then easy to find that, within the discussed  measuring process, the predicted errors of oscillator energy are characterized by the following indicators
\begin{equation} \label{eq:63} 
\mathcal{PEI} \left\{ <H> \right\}=\frac{\omega \left[\hbar ^{2} +\left(\hbar +2m_{0} \omega \gamma ^{2} \right)^{2} \right]}{4\left(\hbar +2m_{0} \omega \gamma ^{2} \right)} -\frac{\hbar \omega }{2} \ne 0 \end{equation} 
\begin{equation} \label{eq:64} 
\mathcal{PEI} \left\{ \mathrm{Var} \left(H\right)\right\} = \left[\frac{\sqrt{2} m_{0} \omega ^{2} \gamma ^{2} \left(\hbar +m_{0} \omega \gamma ^{2} \right)}{\left(\hbar +2m_{0} \omega \gamma ^{2} \right)} \right]^{2} \ne 0 
\end{equation}

\end{document}